\documentstyle[epsf]{mn}

\title[\iras: Modeling the optical spectra from flaring large scale
              jets]
      {\iras:
       Modeling the optical spectra from flaring large scale jets}
\author[Steffen et al. ]
       {W. Steffen,$^{1}$ A.J. Holloway,$^{1}$ A. Pedlar$^{2}$ \\
    $^1$Department of Astronomy, University of Manchester,
          Schuster Laboratory, Oxford Road, Manchester M13 9PL, UK \\
    $^2$Nuffield Radio Astronomy Laboratories, University of Manchester,
       Jodrell Bank, Macclesfield, Cheshire SK11 9DL, UK}

\date{Empty}

\begin{document}

\newcommand{\iras}{IRAS\,04210+0400 }
\newcommand{\kms}{{\rm\,km\,s}$^{-1}$}
\newcommand{\oiii}{[O{\sc iii}]-5007\AA~}
\newcommand{\han}{H$\alpha$+[N{\sc ii}]}
\newcommand{\ha}{H$\alpha$}
\newcommand{\sii}{[S{\sc ii}]}
\newcommand{\etal}{{et al.\,}}

\maketitle

\begin{abstract}
The emission-lines in the active galaxy IRAS\,0421+0400 show a
dramatic ($\sim$\,900\kms) increase in the velocity spread at the
position of radio hot-spots which are located at the beginning of
extended radio lobes.  We study a simple geometric model of an opening
outflow which reproduces the structure found in the long-slit
emission-line spectrum of the hot-spot regions. The
predicted bifurcations in the optical image structure of these regions
is confirmed by deep \oiii\,line-imaging. We propose that this
phenomenon is the result of a jet emerging from the galaxy through the
boundary between the interstellar and intergalactic medium. A similar
model has previously been suggested as an explanation for wide angle
tail radio sources (WAT's). If our model proves to be correct in more
detailed future observations, then IRAS\,0421+0400 provides the unique
possibility to study the flaring jet phenomenon at optical wavelengths.
\end{abstract}


\section{Introduction}
\label{intro.sec}

Jets in FR\,I radio galaxies can flare very abruptly and show very
large opening angles up to $90^{\circ}$ in diffuse lobes or tails
(O'Donoghue \etal 1993).  These structures often bend near the flaring
point. Norman \etal (1988) and Loken \etal (1995) have modelled this
phenomenon in wide angle tail radio galaxies (WAT) in terms of a
supersonic jet passing through a shock in the ambient gas where the
jet flow becomes subsonic. The jet is then disrupted and entrains
external gas, which becomes turbulent with the formation of large and
small scale eddies. Such a shock in the ambient medium could be due to
a supersonic galactic wind moving into the surrounding intergalactic
medium.

In order to obtain direct kinematic information, Owen \etal (1990)
conducted a search for optical line emission from the flaring regions
in WAT sources but found no significant emission from the 5 objects
they studied.  In Holloway \etal (1996, Paper~I) we first proposed
that a phenomenon similar to the WATs could apply to \iras for which
optical information is available in the flaring region (Beichman \etal
1985; Hill \etal 1988; Holloway \etal 1996).  If our interpretation is
correct, then IRAS\,0421+0400\ is an important test object allowing us
to study such a transition region in the optical regime, thereby
providing important kinematic information from spatially resolved
spectra.

The active galaxy IRAS\,0421+0400 was discovered on scans of the
Infrared Astronomical Satellite (IRAS) by (Soifer \etal 1984).
Beichman \etal (1985) identified the IRAS source with a spiral galaxy
at a red-shift of z=0.046. Their spectroscopic work revealed a
Seyfert-2-type emission-line nucleus of the 16.3 magnitude (R-band)
object. First observations with the Very Large Array (VLA) showed an
unusual radio structure, consisting mainly of large symmetrically bent
lobes ($\sim$\,25\,kpc), which seemed to extend the spiral structure
beyond the optically detected image (Beichman \etal 1985). Hill \etal
(1988) reported the discovery of a central double radio source
($\sim$\,1\arcsec separation) and that the lobes start at hot-spots
which are coincident with anomalous spectral features (Steffen \etal
1996a). They found that at the position of the hot-spot the spectral
lines consisted mainly of an asymmetric narrow (fwhm\,=\,160\kms) and
a broad (fwhm\,=\,730\kms) component.  They argued that the extended
spiral features found in this galaxy are the photoionized remnants of
earlier radio jet activity. Steffen \etal (1996b) examined this model
in some detailed and concluded that it was not consistent with the
observations and suggested an alternative model based on the bending
of the jets by the interaction with the rotating interstellar medium.

In Paper~I we identified
broad red and blue shifted velocity components (north and south of the
continuum peak, respectively) with similar spatial separation as the
central double radio source. We spatially resolved the broad spectral
features near the radio hot-spots and presented first results from a
model which explains the spectra as a consequence of an opening cone
starting at the radio hot-spots.  As pointed out in Paper I, the origin
of these structures seems to be related to the flaring of the
radio structure into extended lobes.

Hubble Space Telescope (HST) observations of \iras presented by
Capetti \etal (1996) spatially resolve the central region into a
complex structure with discrete emission-line clouds. There are three
particularly bright knots, confirming the triple structure deduced
in Paper~I from spectroscopic observations. Capetti \etal
(1996) also find filamentary high ionization line emission (coincident
with the previously known spiral features) which bifurcates close to
the southern radio hot-spot.  This strongly supports the model proposed in
Paper~I involving a flaring jet at this position. 

In the present paper, we give a more detailed account of the model
first proposed in Paper I. In Section \ref{model.sec} we introduce the
model, discussing the dependence of the spectral line-shape on the
orientation of the outflow with respect to the observer and the
direction of the spectrometer slit.  In Section \ref{obs.sec} we
compare our model with the observations, which we summarize at the
beginning. In the same section a new deep emission-line image is
presented.  We discuss and summarize our results in Sections
\ref{discuss.sec} and \ref{sum.sec}, respectively.

\noindent

\begin{figure}
\centering
\mbox{\epsfclipon\epsfysize=2.8in\epsfbox[0 0 430 376]{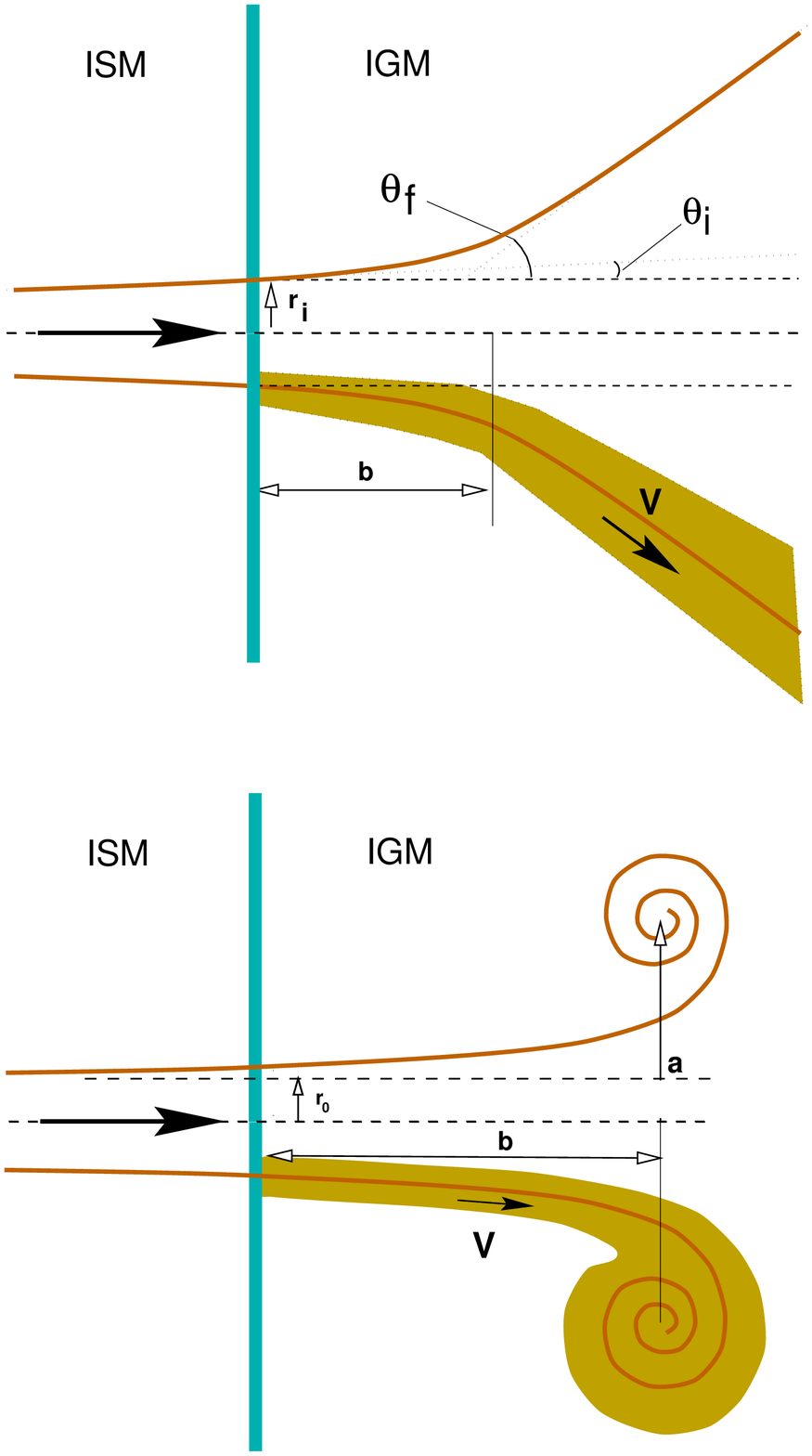}}
\caption{
In this line drawing the geometric model parameters of the flow are
illustrated. They are the initial radius of the circular cross section
$r_0$, the distance $b$ of the eddy from the transition region, where
the emission is assumed to start, and the distance $r_0+a$ from the axis.
}
\label{outfl.fig}
\end{figure}

\section{The model} 
\label{model.sec}

In this section we describe a simple model which is intended to
provide some insight into the existent gross emissivity and flow
distribution in the emission-line gas associated with the extended
radio emission in \iras. We model the long-slit spectra using a
simple parameterized description of the emissivity and the velocity
field of the ionized gas flow. In our model we regard the extended
emission-line source around the radio hot-spots as a collimated
outflow which flares when passing through the boundary between the ISM
and the IGM (see Figure \ref{outfl.fig}).

First we give the details about the parameterization used in this
model.  We then discuss the general properties of the predicted images
and spectra as a function of orientation in space. After discussing the
observational constraints, we present a simple axisymmetric and an
improved non-axisymmetric solution which match the observations in some
details.
\subsection{Parameterization of the model}
Hydrodynamic simulations (Loken \etal 1995) suggest that large-scale
eddies often develop in the flaring regions of jets crossing an
external shock and that they can approximately be described by a
circular or spiraling sheet with flow velocities comparable to the
boundaries of the flaring jet region.  For simplicity we parameterize
this flow as a hyperbolic spiral given by
\begin{eqnarray}
z'_{(s)}  &=& - \frac{a}{s}   \cos(s) 
             + \frac{a}{s_0} \cos(s_0) \\
r_{(s)}  &=& \{r_0 + a [1-\sin(s)/s]\}/\epsilon 
\label{radius_eq} \\
\epsilon &=& [1-e \cos(\phi)]/(1-e) .
\label{epsilon_eq}
\end{eqnarray}
Here ($z'$,$r$,$\phi$) are cylindrical coordinates, and $s$ is a
normalized position parameter along the spiral (with
$s_0=-b/a+\sqrt{2+(b/a)^2}$).  The parameter $r_0$ is the initial
radial distance from the axis, while $\epsilon$ allows for an elliptic
cross section of eccentricity $e$ as a simple approximation to
non-axisymmetry. The distance between the centre of the eddy and the
axis is given by $r_0+a$. The parameters $a$ and $b$ are as indicated
in Figure \ref{outfl.fig}. The orientation of the outflow in space is
determined by the Euler angles ($\Theta$,$\Phi$,$\Psi$) with the
conventions given in Goldstein\,(1980).

The calculations are performed on a regular 3-dimensional cartesian
grid ($x,y,z$), with the $z$-axis along the line of sight. The
$z$-axis of the cartesian coordinate system coincides with the
$z'$-axis of the cylindrical system used for the representation of the
outflow in Equations \ref{radius_eq} \& \ref{epsilon_eq} when the
Euler-angle $\Theta = 0$.  Typical array sizes of the calculations
presented here are 80-100 pixels ($\approx$ 10 pixels per arcsec) along
a side of the cubic domain. Since we are considering mainly forbidden
lines (\oiii \& [N{\sc II}]\,6584-\AA), the emission is assumed to be
optically thin. The image intensity is therefore calculated by adding
the emissivity in each pixel along the line of sight.
\begin{figure*}
\centering
\mbox{\epsfxsize=6.25in\epsfbox[20 0 511 404]{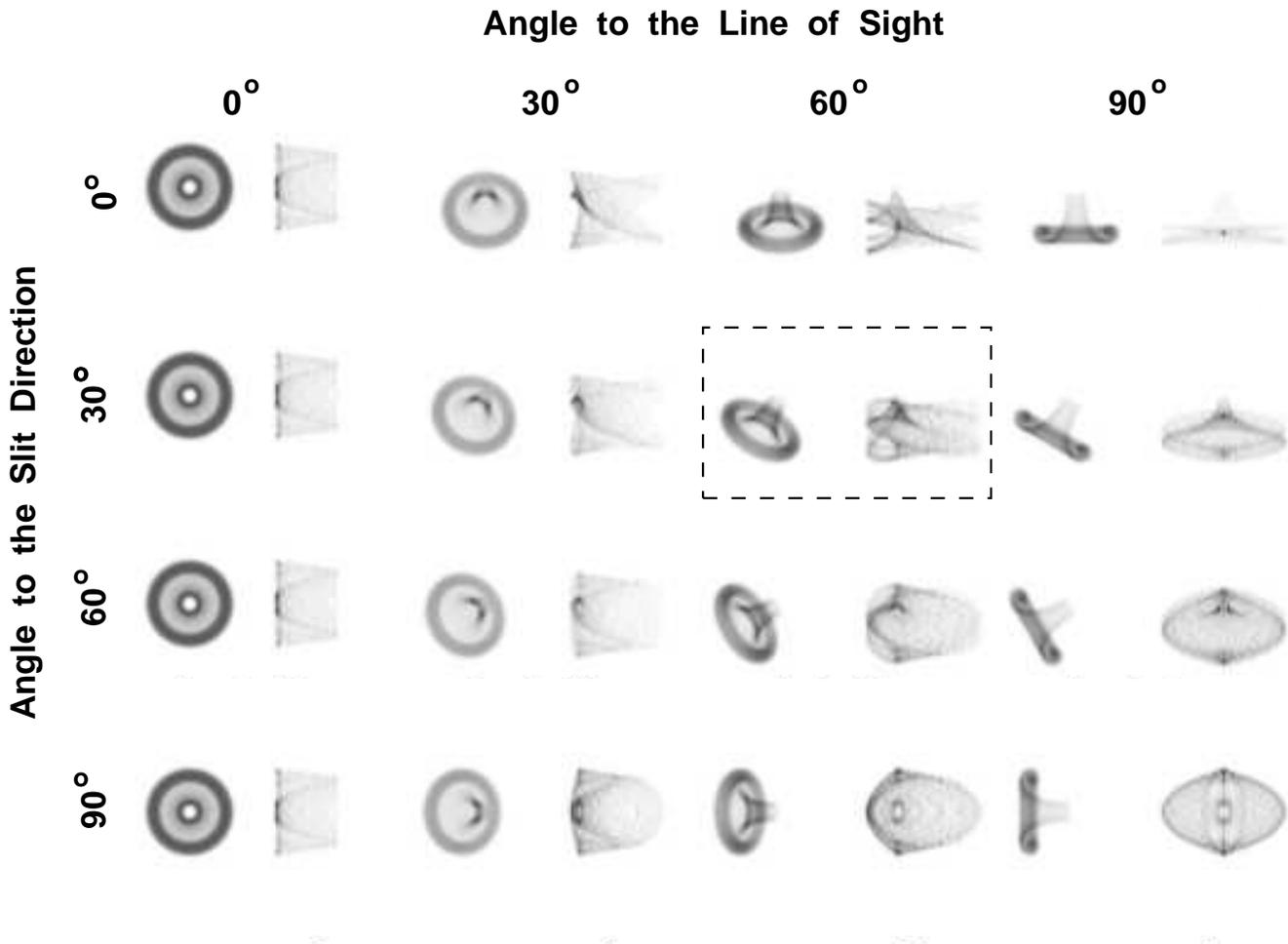}}
\caption{
A series of simulations of an opening outflow 
is shown varying the orientation of the axis with
respect to the line of sight (left to right) and  the
slit orientation. The slit runs from top to bottom and covers the
whole emission. Represented are pairs of images (left) and long-slit
spectra (right). The simulation marked with a dashed box is shown in
more detail in the next figure. In this series it is the closest
match to the observations at the southern hot-spot region in
IRAS\,0421+0400, and it was used as a starting point for more detailed
simulations which are compared to the observations below.
}
\label{mosaic.fig}
\end{figure*}
\begin{figure}
\centering
\mbox{\epsfxsize=3.25in\epsfbox[0 0 511 404]{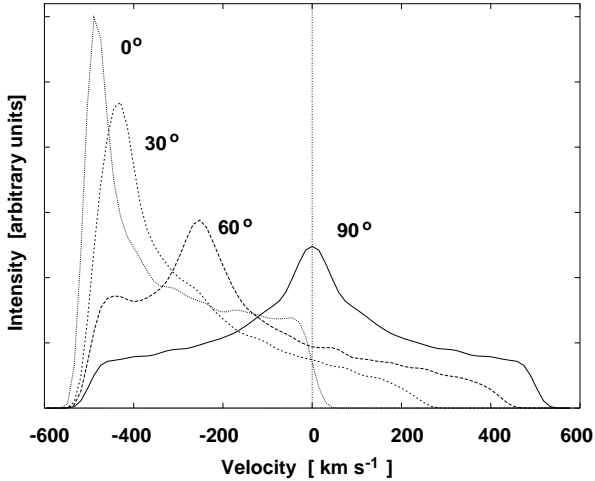}}
\caption{
Integrated spectra are plotted which correspond to the long-slit
spectra from the previous figure. Each of the four spectra represents
all the long-slit spectra for a particular angle to the line of
sight. The four integrated spectra are distinguished by the angle to
the direction with respect to the slit, which is given beside the
curves.}
\label{integ.fig}
\end{figure}

\begin{figure}
\centering
\mbox{\epsfxsize=3.3in\epsfbox[ -10 -10 827 298]{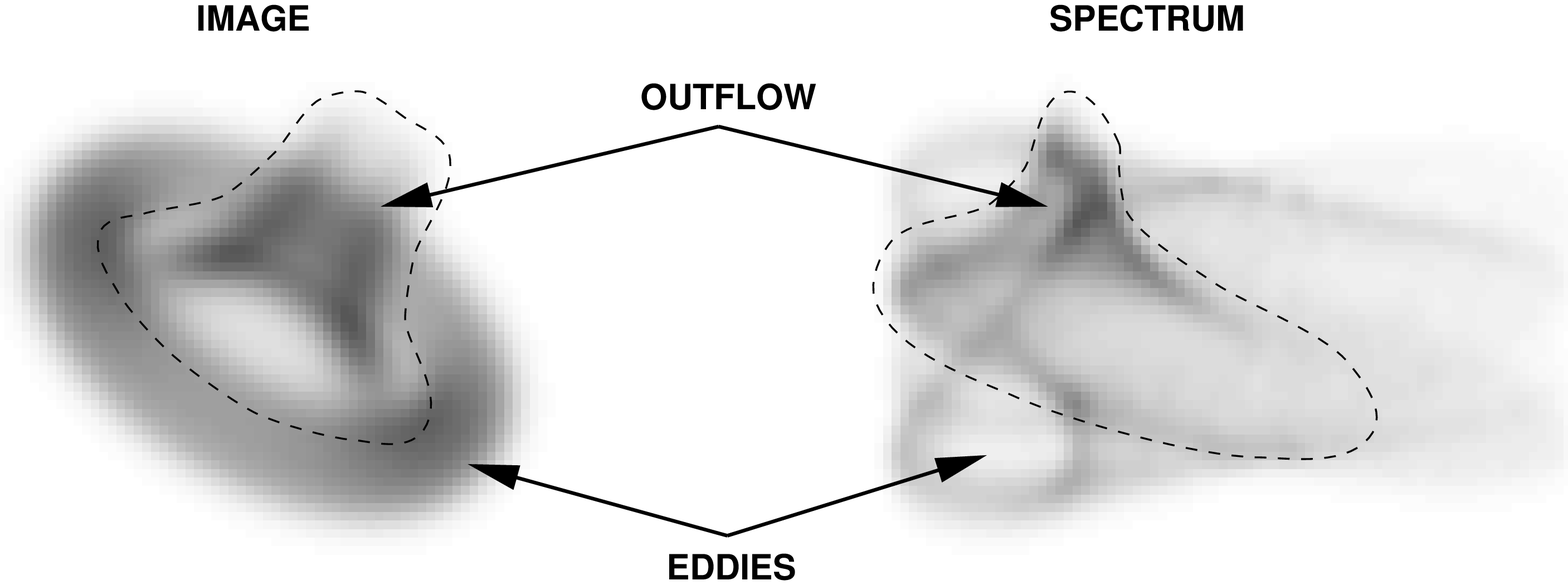}}
\caption{
The simulation marked with a dashed line in the previous figure is
shown in more detail. The out-flowing material, roughly marked with a
dashed line in the image simulation (left), basically produces the
marked region in the long-slit spectrum (right).  Everything outside
this region (and some background contribution to the inside region) is
originated in the back-flow and eddy.  }
\label{detail.fig}
\end{figure}

\subsection{Dependence of the spectrum on orientation}
\label{orient.sec}
The long-slit line profile is most dependent on the orientation of the
outflow with respect to the observer's line of sight and the
orientation of the spectrometer slit. This is true for full
slit-coverage of the outflow. If the slit does not cover the whole
structure, the resulting spectrum is also strongly dependent on the
slit position. This dependence is discussed in more detail in Section
\ref{cover.sec}. The detailed shape of the outflow and the change in
emissivity as a function of distance from the starting point (as, for
instance, a slow decrease in emissivity compared to a constant
emissivity) only influence the details of the simulated spectrum and
not the gross features. The optical emission is rather sensitive to
temperature and density of the flow, and these change considerably
where the outflow turns back and becomes turbulent (Loken \etal
1995). It can therefore be expected that optical emission drops
drastically near this point, rendering the eddies very faint or
undetectable.  Therefore, we assume a constant emissivity and velocity
along the flow line and a Gaussian transverse emissivity distribution
of FWHM$=0.5 r_0$, with a cut-off at position $s=s_f$ along the
spiral (which we expect to be near the turning point).

In Figure \ref{mosaic.fig} we show a series of results for
axisymmetric outflows varying the angle between the axis and the line
of sight (left to right) and orientation with respect to the direction
of the spectrometer slit (which covers the whole structure). A
nonlinear grey-scale is used to emphasize the background emission in
the spectra which arises mostly from the eddies. The outflows shown
here are directed towards the observer. The emission is therefore
mainly blue-shifted. Since we are considering only optically thin
emission, corresponding orientations away from the observer produce
symmetrically red-shifted spectra and are omitted.  For most
orientations the opening section of the outflow produces a bright,
asymmetric, and opening feature in the spectrum which is often
`$\Lambda$'-shaped (or `V'-shaped, if the outflow is oriented in
opposite direction). The eddies generally result in a diffuse
distribution around the emission arising from the outflow.  The
simulation marked with a dashed box in the mosaic (Figure
\ref{mosaic.fig}) is shown in detail in Figure
\ref{detail.fig} with the outflowing region emphasized in the image
and the spectrum. It shows the back-flow can produce loop-like
structures.  

In Figure~\ref{integ.fig} we display the integrated spectra
corresponding to the simulations in the mosaic. Note that the spectra in
each column of Figure~\ref{mosaic.fig} have the same integrated
spectra and need to be spatially resolved to be distinguishable from
each other. The integrated spectra consist of blue-shifted peaks with
broad wings mainly from the eddies.

\begin{table}
\caption{The geometric parameters and the velocity $v$ of the simulations, 
which we compare with the observations are given. The first two rows
are axisymmetric simulations of the northern (N) and southern
(S). ($\Theta$,$\Phi$,$\Psi$) are the Euler angles giving the
orientation in space with the conventions of Goldstein
(1980). Herein $\Theta$ is to the angle with respect to the line of
sight and $\Psi$ corresponds the angle with respect to the slit
orientation. The azimuthal orientation of the outflow on the axis is
given by $\Phi$. The point at which the emission ceases along the
outflow is $s_f$. }
\begin{center}
\begin{tabular}{llllllllll}
\hline
e   &   & $v$  &   $\Theta$   & $\Phi$  & $\Psi$  & $b/a$  &  $s_f$  \\
\hline
0 & (N)  & 350  & $100^{\circ}$  & $  -  $   & $35^{\circ}$  & 0.45  &  1.7 \\
0 & (S)  & 500  & $-105^{\circ}$ & $  -  $   & $30^{\circ}$  & 0.8   &  2.5 \\
0.7 & (S)& 500  & $-75^{\circ}$  & $-35^{\circ}$ & $35^{\circ}$ & 0.5 & 3.2 \\
\hline
\end{tabular}
\end{center}
\end{table}
\begin{figure*}
\centering
\mbox{\epsfxsize=6in\epsfbox[0 0 511 458]{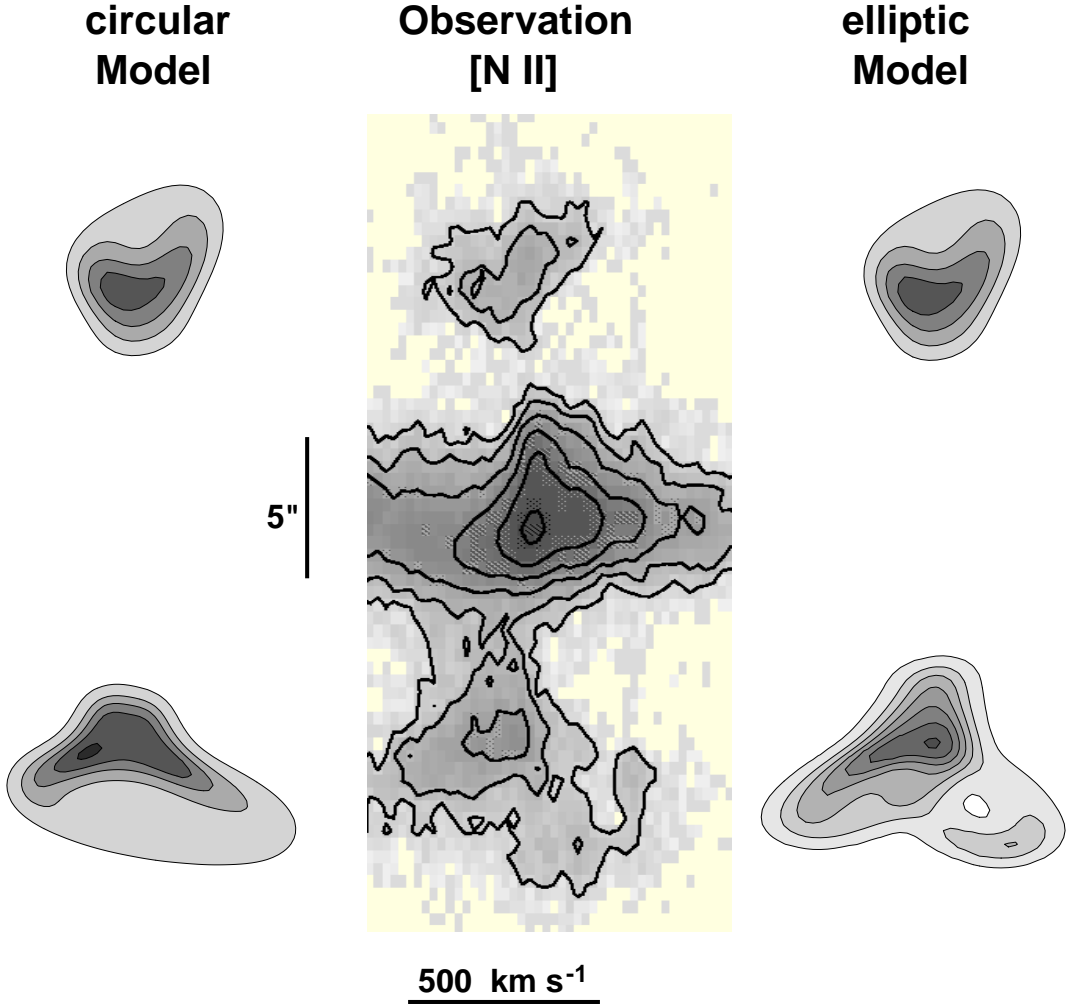}}
\caption{
The observed [N{\sc II}]\,6584-\AA long-slit spectrum (middle) is compared to
our model. On the left the model with the fully circular cross section
is represented. On the right of the observations, the results for
model variant with an elliptic cross section in the southern outflow 
are displayed. Observe the flaring of the spectra at
approx. 5\,arcsec to the north and south of the centre. }
\label{spectra.fig}
\end{figure*}

\begin{figure}
\centering
\mbox{\epsfxsize=3.25in\epsfbox[0 0 394 559]{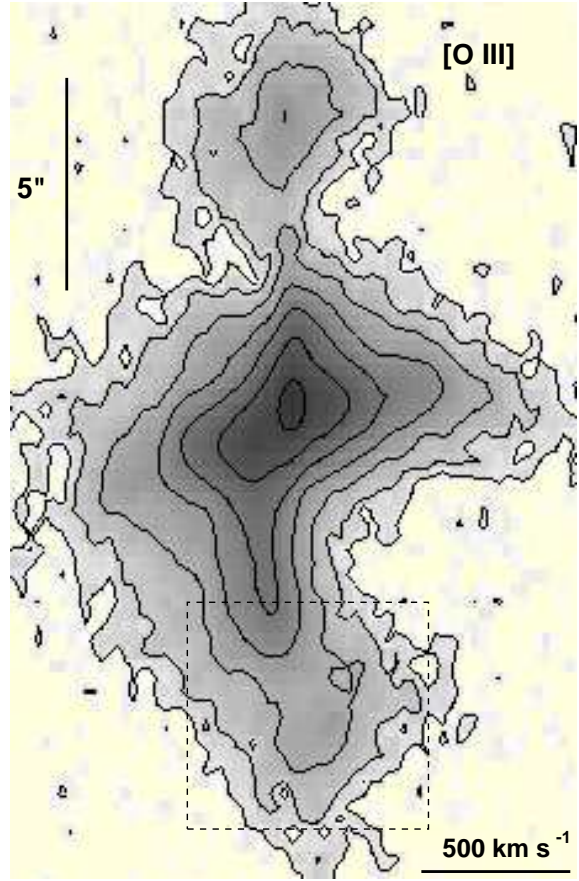}}
\caption{The long-slit profile of the \oiii line (from Paper~I). 
The dashed box marks the southern hot-spot region.}
\label{oxy3.fig}
\end{figure}

\begin{figure*}
\centering
\mbox{\epsfxsize=6in\epsfbox[0 0 511 488]{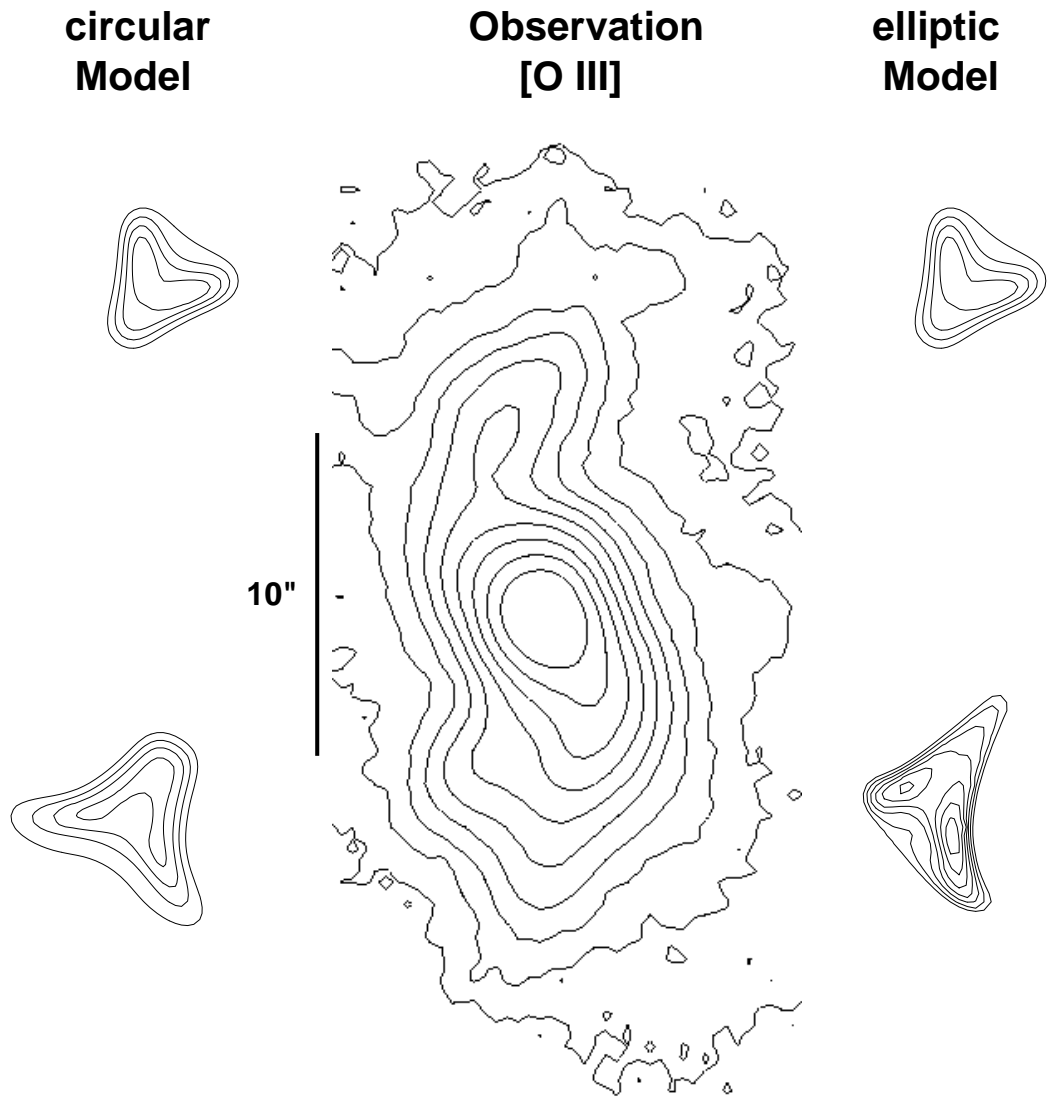}}
\caption{
The middle section shows the observed \oiii emission-line image of
\iras. Note the bifurcations to the north and south
of the centre, which we identify with the opening outflows.  The
predicted model images of the flaring regions in the case of an
axisymmetric outflow are shown on the left side of the observations.
On right, the simulation where the southern hot-spot region has an
elliptic cross section is represented. Contour levels of the
observation are 0.5,\,1,\,1.5,\,3,\,6,\,10,\,15,\,25,\,40,\,60\,\% of
the peak brightness.  For clarity, the contours of the simulated
images have equally spaced contours.}
\label{observe.fig}
\end{figure*}

\section{Observations}
\label{obs.sec}

\subsection{Long-slit spectra}
\label{spectra.sec}
The long-slit observations to which we apply our model where presented
in Paper~I, therefore suffice to summarize the main properties which
are important for our model.  The basic structure can be appreciated
best in the [N{\sc ii}] 6584-\AA\ line, which we reproduce in Figure
\ref{spectra.fig}. The northern hot-spot at around 5\arcsec\
separation from the centre shows a `V'-shaped structure, where the
red-shifted arm is stronger and more extended. 

The southern hot-spot shows a very similar, `$\Lambda$'-structure, which
is better defined: most of the emission spreads into a blue wing. At a
lower brightness level, emission is seen in a red wing. The
observation is suggestive of two or more discrete components or
possibly a ring-like structure in the velocity-space map. A
considerably different picture is seen in the \oiii\ line (reproduced
in Figure \ref{oxy3.fig}). Here the `$\Lambda$'-shape 
is not obvious. Instead, all or most emission seems to be in one of the
red-shifted components also seen in the [N{\sc ii}] 6584-\AA\ line.

As mentioned in Paper~I, there are three simple possible kinematic
interpretations of such `$\Lambda$'-shaped structures in the velocity
map. First, they may arise from real acceleration of the
gas. Alternatively, the gas might have a certain high velocity when
the line emission starts and only the direction of the velocity vector
changes along its trajectory. Naturally, the third possibility is a
combination of both effects. The bifurcations found in the new optical
line images presented in the next section provide new evidence for our
model of an opening out-flow, in which the flaring of the spectra in
the hot-spots is dominated by geometric effects (i.e. the flow
direction changes along the streamlines).

\subsection{New emission-line imaging}
\label{image.sec}

\begin{figure*}
\centering
\mbox{\epsfclipon\epsfxsize=4in\epsfbox[10 20 568 566]{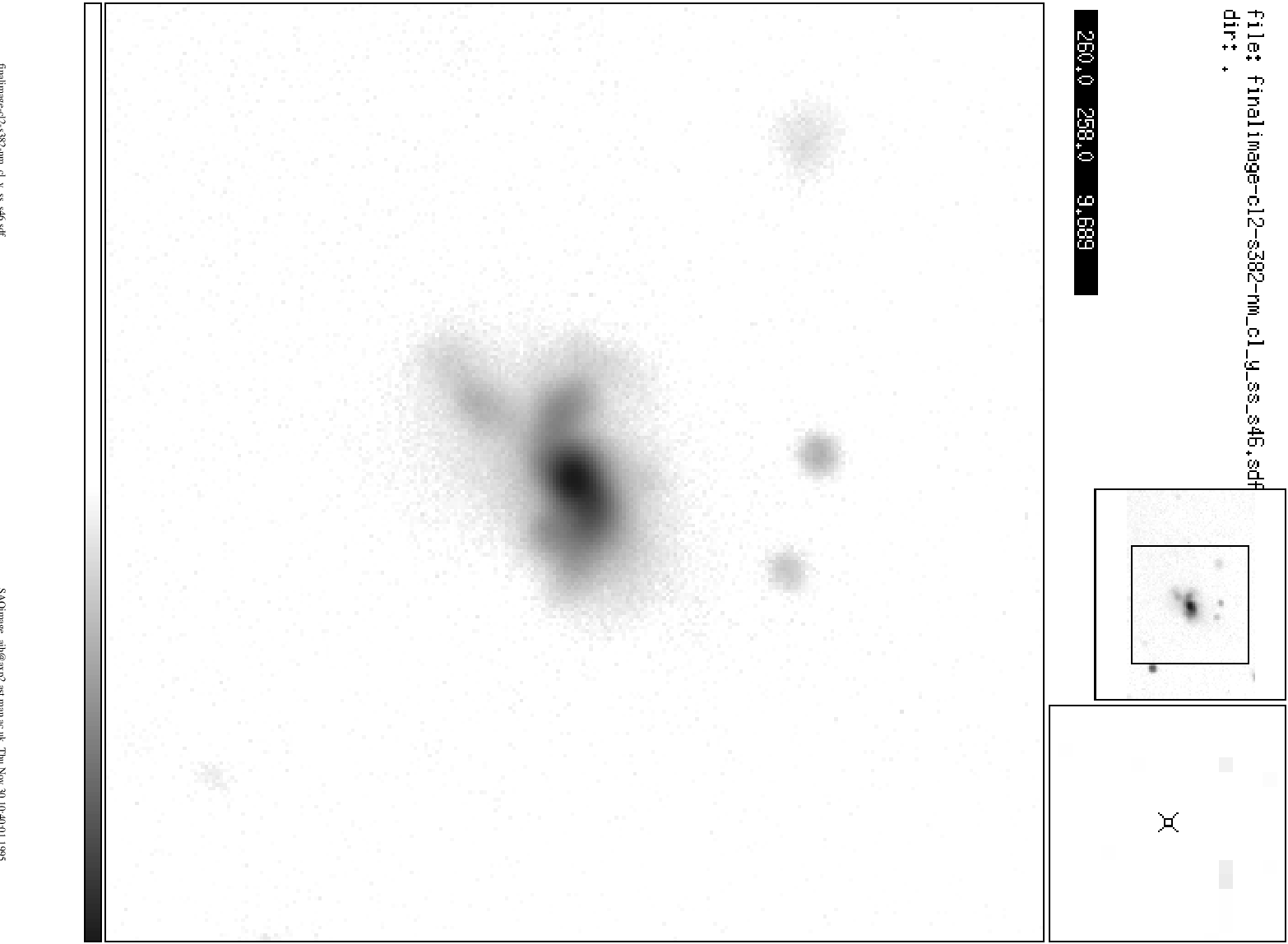}}
\mbox{\epsfclipon\epsfxsize=4in\epsfbox[10 20 568 566]{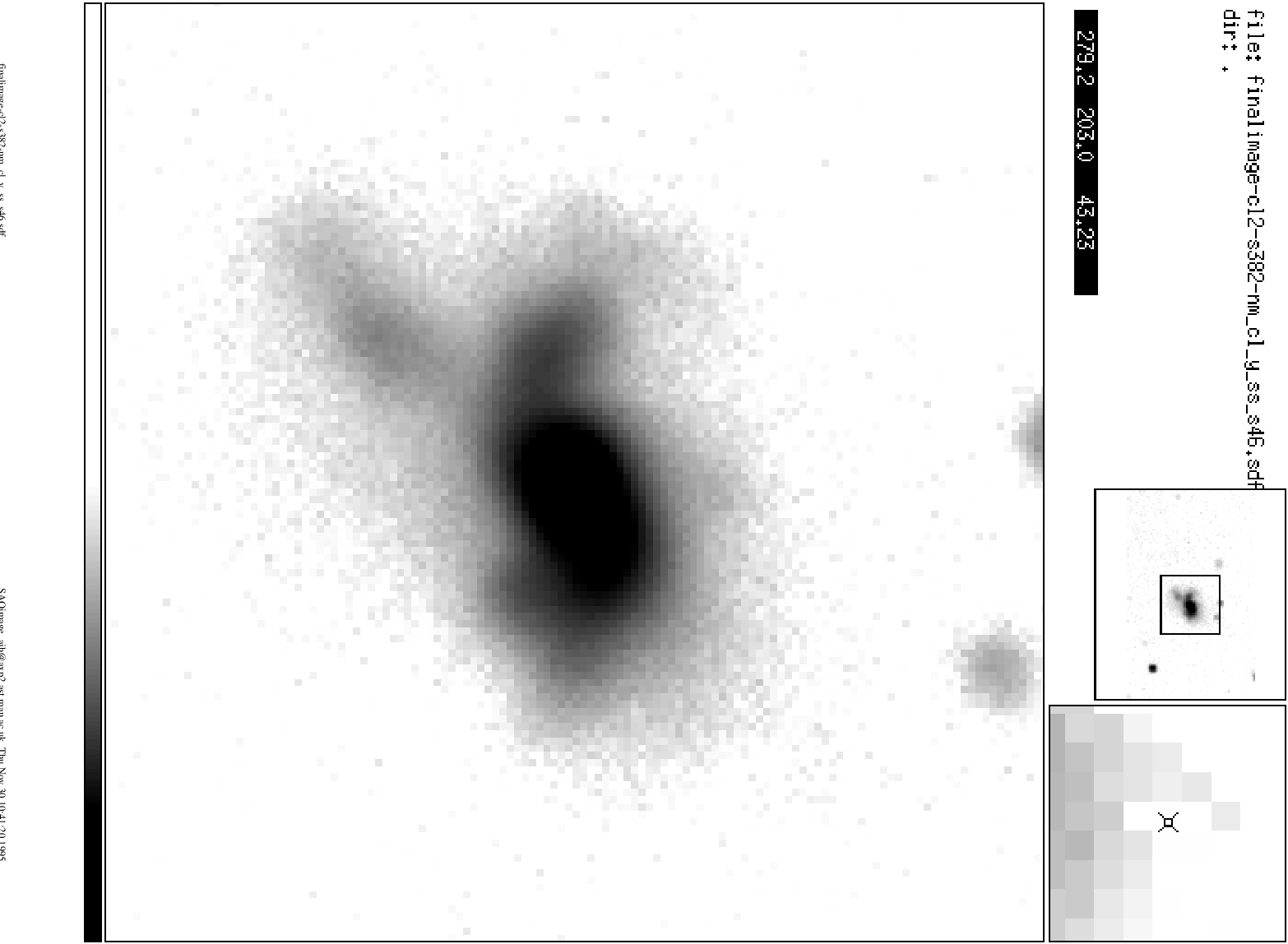}}
\caption{A grey-scale representation of a deep \oiii\ image taken
at the Anglo-Australian Telescope is shown in the top panel (a), which
emphasizes low brightness features at the edge of \iras. Note the
eak, but clearly detected, bifurcation at the northern edge (north is
towards the top). It also shows the near environment which is composed
by a small companion galaxy at $\sim 11$\arcsec and other barely
resolved objects. Several unresolved foreground stars are also seen.
The bottom panel (b) shows \iras only with its companion. The grey-scale
was chosen such that the emission-line spiral structure is emphasized. }
\label{plate.fig}
\end{figure*}

In Figure~\ref{plate.fig} (Plate~1) we present a new [O\,{\sc iii}]
5007-\AA\ emission-line image of \iras. It is a composite of three
images of 30 minutes exposure time each, obtained at the
Anglo-Australian Telescope (AAT) with an [O\,{\sc iii}] 5007-\AA\
filter (bandwidth 70\AA).  We show two versions of the same data.  At
the top of Figure~\ref{plate.fig} the galaxy is shown including its
near environment with the grey-scale chosen such that the fainter
extended emission is emphasized.  In the bottom picture the bright
central spiral structure is emphasized. The seeing varied from 1 to
2\,arcsec between the individual exposures.  The image confirms the
central structure observed previously by Beichman \etal (1985). It
shows the bifurcation of the `spiral arm' in the south, which is also
present in the HST-image obtained by Capetti \etal (1996).  However,
our deep ground-based image clearly reveals a faint bifurcation of the
extended emission-line spiral arm beyond the northern hot-spot
region. This observation strongly supports our interpretation of the
flaring of the emission-line width in these regions as an opening
out-flow of emission-line gas. This northern `V' is much fainter than
the strong bifurcated southern `arm'.

\subsection{Constraints}
\label{constraints.sec}
Before we show the results of detailed models of the observations, we
summarize the constraints which can be used as an input for the
modeling process. From the new \oiii\ image, the projected direction
(angle $\Theta$) to the axis of the out-flow can be estimated from the
orientation of the bifurcations. The result is consistent with the
orientation of the radio lobes. We find that the approximate position
angle of the bifurcations in the north is $-35^\circ$ and the south is
$140^\circ$. Other constraints are the spatial extension and radial
velocity width of the emission in the hot-spot regions, the asymmetry
of the emission in the spectra, and the position of the maximum in the
spectra, but which are more difficult to quantify. 

The large parameter space has been explored by performing large
numbers of intermediate resolution simulations ($\approx$ 50 pixels
along each axis) varying 2 parameters at a time. The parameter grid
was chosen fine enough to ensure that only small changes were found
between individual frames. The resulting series of images and spectra
were inspected, and promising parameter sets were selected and
refined. Although a few sets of parameters have been found which
reproduce the observations partially, the solutions discussed in the
following section give by far the best overall match between the
simulation and the observations.

For direct comparison, the simulated images and spectra have been convolved 
with Gaussian point spread functions with FWHM similar to the resolutions of 
the observations (with 100 velocity channels). 

\subsection{Comparison between model and observations}
In Figures \ref{spectra.fig} and \ref{observe.fig} we compare the
observed long-slit spectrum of the [N{\sc II}] 6584-\AA\ line and our
new \oiii\ line-image, respectively,
with our model. The simulation shown on the left is an axisymmetric
calculation for both, the northern and the southern hot-spot region.
The important parameters are given in rows 1 \& 2 of Table~1. Note
that the flaring and the rough brightness distribution in the spectrum
are reproduced. However, one does also note that the red wing in the
southern hot-spot is of similar strength as the blue wing. Also, the
maximum emission is within the blue wing. These are general properties
the axisymmetric simulations which fulfill the geometric constraints
given in Section
\ref{constraints.sec}. They are in conflict with the observation.

These problems are solved in the simulation on the right side of Figures
\ref{spectra.fig} \& \ref{observe.fig}. Here
the northern model is the same as before, but an elliptic cross
section according to Equation \ref{epsilon_eq} was used with an
eccentricity of $e=0.7$. Similarly, the radial velocity component is
now $v_{r(\phi)} = v_{r0} \epsilon_{(\phi)}$. Note that the velocity
is still constant along each flow line. In this variant of the model,
the low brightness red-shifted section of the spectrum (with the shape
of a loop) is produced close to the cut-off of the emissivity at
$s_f$. Here the cut-off occurs where the out-flow starts to flow back,
while in the axisymmetric case the emission cuts off before any
back-flow or eddies develop. Note that extending the emission as far as
in the elliptic case does not improve the result, but becomes
completely inconsistent with the observation. The loop in the elliptic
simulation is similar to the loop in Figure
\ref{detail.fig}, only that the red-shifted emission is now
blue-shifted and vice-versa (since it is pointing away from the
observers instead of facing it).

Another improvement of the elliptic model over the axisymmetric case
is seen from Table~1. In the axisymmetric simulation, the northern and
the southern out-flow necessarily have to be pointing towards the
observer, i.e. they are not co-linear (the angle between their axes is
$\sim 155^\circ$). In contrast, in the non-axisymmetric simulation the
axes form an angle of $\sim 175^\circ$, which is consistent with the
projected symmetry suggested by the radio lobes.

\subsection{Slit coverage and the \oiii line}
\label{cover.sec}

In this section we consider the effect of slit coverage of the out-flow
on the appearance of the spectrum. The simulations in the previous
section assumed full coverage of the out-flow. We note however that the
size of the slit in the observations was 1\arcsec and the images of
the out-flows exceed this size in east-west direction. Another
motivation for investigating reduced slit coverage was a significant
qualitative difference between the southern hot-spot region in the
\oiii\ line and the \han\ complex. We suspect that this could be due to
insufficient slit coverage.  As can be seen in comparing Figures
\ref{spectra.fig} and \ref{oxy3.fig}, the southern
\oiii\ emission does not show the bright blue and red-shifted wings
present in the \han and the \sii\ lines at $\sim\,$5\arcsec from the
galaxy centre. A rather discrete red-shifted component is
present at $\sim\,$7\arcsec core distance (see Figure \ref{oxy3.fig}).
\begin{figure*}
\centering
\mbox{\epsfxsize=6in\epsfbox[0 0 403 279]{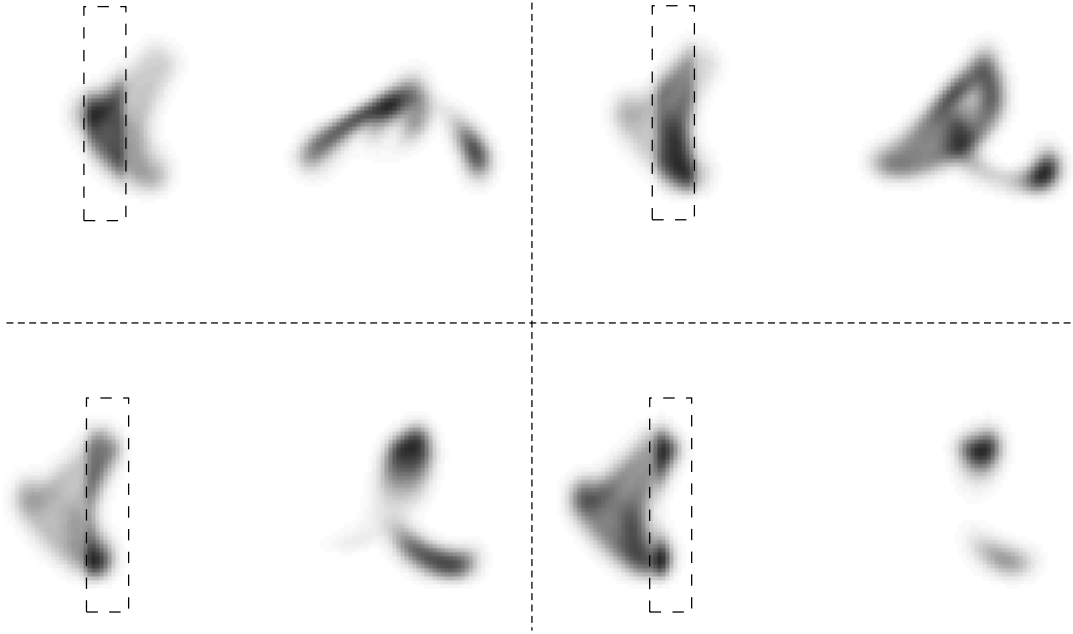}}
\caption{The southern out-flow with elliptic cross section is shown
for different slit coverages.  As in the previous Figures, the pairs
of image (left) and long-slit spectrum (right) are shown. The section
included in the spectrum is marked by a dashed box and with the
intensity enhanced by a factor of three in the images.}
\label{step_mark.fig}
\end{figure*}

We consider two possible explanations for this difference. The first
of these concerns the uncertainty in the slit coverage of the relevant
structures. We note that the \han\ and \sii\ (not all shown here)
results are from a single observations, while the \oiii spectrum was
obtained on another day and different seeing conditions.  The second
possibility is an intrinsic difference in the distribution, possibly
caused by the different expansion velocities on a non-axisymmetric
opening cone, which could result in different emission properties.
 
We performed a set of simulations involving a slit which only
partially covers the image structures and step it from east to
west. We used the non-axisymmetric simulation which compares best with
the observations (Figure~\ref{spectra.fig}).  In
Figure~\ref{step_mark.fig} the area covered by the slit is marked by a
box and the brightness is enhanced by a factor of 3 compared to the
uncovered portion in the image. As expected the observed spectrum
depends on the slit coverage. However, we also notice that, with the
slit covering the central section of the out-flow, the upper
panels contain all the essential spectral features of the observed
\han\ lines as shown in the simulation with full coverage in
Figure~\ref{spectra.fig}. This justifies the reduction of the number
of free parameters by choosing full coverage in the simulation of
observations with adequate slit positioning.

We also find that inadequate slit coverage could be the
reason for the observed difference between the \oiii\ line and the
\han\ complex.  The simulation on the lower right, with the slit
covering only small image sections, compares well with the
\oiii\ spectrum. The position of the \oiii feature is reproduced in 
position and spectral shift.  From these considerations we conclude,
that the observed difference between the observed
\oiii\ and the corresponding \han\ emission can be explained by an
off-set in the positioning of the slit. It probably adequately covers
the northern hot-spot region, but only partially intersects the
southern counterpart. The shift of the slit position corresponds to
$\sim 1.5$\arcsec and is within the uncertainty of $\sim 2$\arcsec
from the positioning of the spectrometer slit during the observations.

This result provides the simplest explanation of the
difference. However, as discussed in Section \ref{discuss.sec}, it
cannot rule out an intrinsic origin of the difference.  Note that in
the framework of our model, the \oiii\ component arises close to the
end of the opening cone. A final decision can only be made with
further observations with full spectroscopic coverage and detailed
physical modeling of the suggested boundary layer between the jet
material and the external medium.
 
\section{Discussion}
\label{discuss.sec}

We considered the kinematics in the extended emission-line region of
the active galaxy \iras. The distribution and kinematics of the
emission-line gas seems to be strongly linked to the expansion of extended
radio lobes.  We have presented a simple model which describes the
emission as an opening cone with possible eddies. We calculated model
spectra of such out-flows varying the orientation with respect to the
observer's line of sight and the orientation of the spectrometer slit
(the most important parameters determining the shape of the long-slit
spectra). We find that the main features of the observed long-slit
spectra of the extended emission can be explained in terms of two
axisymmetric out-flows which open very rapidly (over $\sim 1$\,kpc at a
projected distance of $\sim 4.5$\,kpc). 

Both, image and spectrum of the northern hot-spot region are well
described by the axisymmetric model. However, the axisymmetric case is
only of limited use for the southern hot-spot region which shows more
detailed structure. A significant deviation of the axisymmetric model
spectra from the observations of the southern hot-spot region is
eliminated allowing for considerable non-axisymmetry of the out-flow.
Even the simplest possible case, an elliptic cross section provides a
better match to the observations than the axisymmetric model.
Moreover, the predicted image of the elliptic simulation
(Figure~\ref{observe.fig}, right) is in better agreement with our
observations and the high resolution HST-image presented by Capetti
\etal (1996).

From our considerations of the difference between the \oiii\ and the
\han\ spectra of the southern hot-spot region, we find that these
could possibly be accounted for by insufficient slit coverage during
the \oiii\ observation. Therefore, re-observation of the \oiii\ line
with adequate coverage and sensitivity is desirable. If the difference
is confirmed to be intrinsic, it could provide important information
to test a more physical model based on the geometric
approach of this paper.

We suggest that the most plausible process producing the morphology
and the flow pattern represented by our model is a turbulent mixing
layer similar to one described by Cant\'o \& Raga (1991) which could
develop rapidly after the jet crossed a shocked boundary between the
interstellar and the intergalactic medium. However, in contrast to the
conditions in their model, which was applied to a supersonic stellar
jet, in our case we would have a hot internally subsonic flow, which
heavily entrains external gas. This gas is heated and ionized.  The
temperature will rise from the edge towards the axis, allowing
radiation coming from recombination to arise only in a thin sheet
close to the edge of the boundary layer. Additional effects could
arise from transverse shock waves send into the external medium as the
jet material expands transversely to its main flow direction
(Komissarov 1994). At the beginning of the flaring jet region,
shock-induced pressure and density of the external gas can be expected
to be highest. Coupled with photo-ionization from the centre of the
galaxy, this could cause reduced \oiii\ emission compared to the
emission in the region of lower compression further downstream.

The model parameters (Table~1) indicate that the inclination of the
axes of the out-flows with respect to the plane of the sky is very
small ($\sim 10^\circ$). Hill \etal (1988) suggested that the
emission-line spiral structure is due to the residual of the
interaction between the jet and the interstellar medium, which then is
dragged away from the jet path by rotation. Since the spiral features are
observed out to the visible edge of the (disk) galaxy, the jet must have
propagated close to the plane of the galaxy in order to encounter
sufficiently dense material to interact with up to the edge of the
galaxy. In this case the distance measures
along the jet line are largely unaffected by projection effects. If
\iras is a disk galaxy, the almost circular appearance of the
continuum image (Paper~I) suggests that it oriented close to
face-on. Unfortunately the classification of the galaxy is still
unclear. In order to settle this question detailed optical and radio
observations will be necessary to separate the effect of the
emission-line spiral structure from the star light and the paths of
the jets.  The spiral structure and the implications from a model
invoking a jet interacting with the external medium will be studied in
a separate paper (Steffen \etal 1996b).

\section{Conclusions}
\label{sum.sec}
We discussed a simple geometric model of an opening out-flow which
reproduces the structure found in the long-slit emission-line spectrum
of the hot-spot regions in IRAS\,0421+0400. The predicted optical
image structure of these regions is confirmed by deep [O\,{\sc iii}]
line-imaging. The model parameters indicate that the axes of the
out-flows lay very close to the plane of the sky, suggesting that the
projected distances along their line are largely unaffected by
projection. We proposed that a jet crossing a shocked boundary between
the interstellar and intergalactic medium could produce such an
out-flow in a mixing layer between the jet material and the
intergalactic environment. If this model is correct, then
IRAS\,0421+0400 provides a unique object allowing us to study this
phenomenon at optical wavelengths. To date, this has not been possible
for the Wide Angle Tail Radio Sources, for which this mechanism has been
proposed previously. A more detailed study of this, so far unique
source, with as high spatial and spectroscopic resolution and
sensitivity is therefore highly desirable.

\section*{Acknowledgements}
AJH and WS acknowledge the receipt of a PPARC studentship and a PPARC
research associateship respectively. 
We thank R.J.R. Williams and A.C. Raga for useful discussions.

\end{document}